\documentclass[amsmath,amssymb,prl,superscriptaddress,twocolumn,normalem]{revtex4-1}
\usepackage[dvips]{graphicx}
\usepackage{booktabs}
\usepackage[USenglish]{babel}
\usepackage{amsthm,dsfont,url,braket,hyperref,mathtools,commath}
\usepackage[percent]{overpic}
\usepackage{float}
\usepackage{lineno}
\usepackage{amsfonts}
\usepackage{wasysym}
\usepackage{mathrsfs}
\usepackage{yfonts}
\usepackage{bbold}
\usepackage{verbatim}
\usepackage{subfigure}
\usepackage{color}
\usepackage{soul,xcolor}
\usepackage[normalem]{ulem}

\newcommand\stb{\bgroup\markoverwith
{\textcolor{blue}{\rule[.5ex]{2pt}{0.4pt}}}\ULon}

\newcommand{\I}{\mathscr{I}}
\newcommand{\dm}{\hat{\rho}}
\renewcommand{\a}{\alpha}
\renewcommand{\b}{\beta}

\begin{document}

\title{One-photon measurement of two-photon entanglement}
\author{Gabriela Barreto Lemos}
\email{Gabrielabl@if.ufrj.br} \affiliation{Institute for Quantum Optics and Quantum Information,
Austrian Academy of Sciences, Boltzmanngasse 3, Vienna A-1090, Austria.}
\affiliation{Instituto de F\'isica, Universidade Federal do Rio de Janeiro, Av. Athos da Silveira Ramos 149,
Rio de Janeiro, CP: 68528, Brazil. }
\author{Radek Lapkiewicz}
\affiliation{Institute of Experimental Physics, Faculty of Physics, University of Warsaw, Pasteura 5, Warsaw 02-093, Poland  }
\author{Armin Hochrainer}
\affiliation{Institute for Quantum Optics and Quantum Information,
Austrian Academy of Sciences, Boltzmanngasse 3, Vienna A-1090, Austria.}
\affiliation{Vienna Center for Quantum Science and Technology (VCQ), Faculty of Physics,
University of Vienna, Boltzmanngasse 5, Vienna A-1090, Austria.}
\author{Mayukh Lahiri}
\email{mlahiri@okstate.edu}
\affiliation{Department of
Physics, Oklahoma State University, Stillwater, Oklahoma, USA}
\author{Anton Zeilinger}
\email{anton.zeilinger@univie.ac.at}
\affiliation{Institute for Quantum Optics and Quantum Information,
Austrian Academy of Sciences, Boltzmanngasse 3, Vienna A-1090, Austria.}\affiliation{Vienna Center for Quantum Science and Technology (VCQ), Faculty of Physics,
University of Vienna, Boltzmanngasse 5, Vienna A-1090, Austria.}

\begin{abstract}
 Measuring entanglement is an essential step in a wide
range of applied and foundational quantum experiments. When a two-particle quantum state is not pure, standard methods to measure the entanglement require detection of both particles.
We realize a conceptually new method for verifying and measuring entanglement in a class of two-party mixed state, for example,
twin photons. Contrary to the approaches known to date, in
our experiment we verify and measure entanglement in mixed
quantum two-party states by detecting only one subsystem, the
other remains undetected. Only one copy of the two-photon mixed or pure state is used but that state is in a superposition of having been created in two identical sources. We show that information shared
in entangled systems can be accessed through single-particle
interference patterns. Our experiment enables entanglement
characterization even when one of the subsystems cannot be
detected, for example when suitable detectors are not available.
\end{abstract}
\maketitle

 Recent developments in quantum technology necessitate measurement of entanglement in a wide  class of systems.  Bipartite pure state entanglement can always be verified and quantified by performing local measurements on only one particle (subsystem), and ignoring the other particle (subsystem)\cite{walborn2006,Pires2009,Just2013}. However, for mixed states,
known schemes for analyzing 
bipartite entanglement,  for example, testing Bell's inequalities \cite{PhysRevLett.28.938,PhysRevLett.49.91,hensen2015loophole,PhysRevLett.115.250401, shalm2015strong}, quantum state tomography \cite{James2001}, testing entanglement witnesses \cite{PhysRevA.66.062305,PhysRevLett.105.230404, PhysRevA.91.032315,PhysRevLett.113.170402,barbieri2003detection}, measuring entanglement using  multiple copies of the state \cite{Schmid2008,PhysRevA.84.052112,zhang2013direct,
islam2015}, all 
rely on the detection of both subsystems. 
Whether entanglement of a bipartite mixed state can be verified
by performing a measurement on only one subsystem is an open question. 
\par
We address this question and demonstrate that it is possible to verify entanglement in a class of bipartite mixed states, encoded in the polarization of two photons,  by detecting only one subsystem and ignoring the other.
We perform an experiment in which the single-photon interference
patterns generated by emissions from two identical twin photon sources contain the complete
information about entanglement in a two-photon mixed state.  Only one photon pair is produced in each detection run, therefore the protocol does not require more than one copy of the same state, instead it uses one copy of the state in a superposition. For certain choices of measurement bases, single-photon interference is possible only when the photon pair is entangled in polarization. The
interference visibility is linearly proportional to the concurrence,
a widely used entanglement measure for qubits.
In fact, even though each photon from a completely mixed (separable) two photon polarization state is described by the same unpolarized state as each photon from a maximally entangled two-photon polarization state, these two scenarios can be distinguished in our experiment without coincidence detection or any post-selection. This experiment was performed simultaneously with the corresponding theoretical study \cite{pol-ent-theory}.
\par

\begin{figure}\centering
  \includegraphics[width=0.95\linewidth]{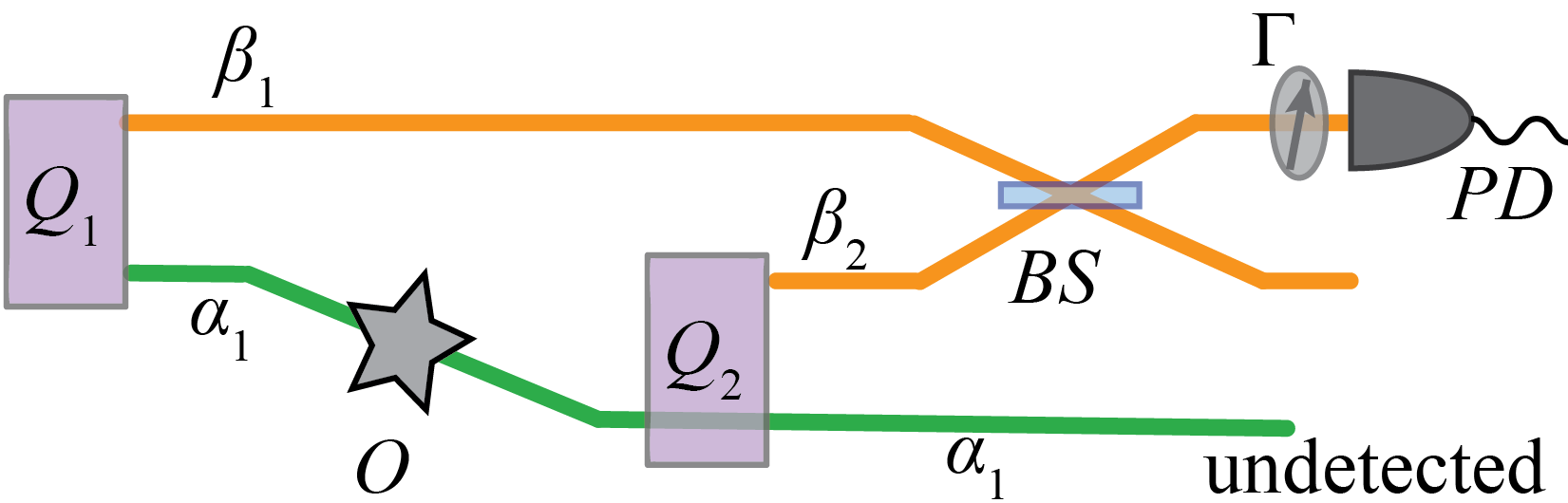}
\caption{Entanglement verification scheme. Two identical sources, $Q_1$ and $Q_2$, individually generate the same two-photon state ($\dm$). Source $Q_1$ can emit a photon pair ($\alpha$, $\beta$) into propagation modes $\alpha_{1}$ and $\beta_{1}$. Source $Q_2$ is restricted to emit photon $\a$ also in the mode $\a_1$. Photon $\alpha$, which is never detected, interacts with a device, $O$, between $Q_1$ and $Q_2$. Source $Q_2$ can emit photon $\b$ in propagation mode $\b_2$. Modes $\beta_1$ and $\beta_2$ are combined by a beamsplitter ($BS$) and an output of $BS$ is collected by a photo-detector ($PD$). Another device ($\Gamma$), placed before $PD$, allows us to choose the measurement basis. Sources $Q_1$ and $Q_2$ never emit simultaneously and stimulated emission in $Q_2$ due to the insertion of $\alpha_1$ mode is negligible.  When it is impossible to know the source of a detected photon, single-photon interference is observed at $PD$. Information about the entanglement is retrieved from the single-photon interference patterns. In the supplementary material we present an alternative setup that does not require two identical sources because the laser passes twice the same pair of crystals.}
\label{fig:scheme-illus}
\end{figure}
We employ two identical sources, $Q_1$ and $Q_2$ (Fig. \ref{fig:scheme-illus}), each of which can generate the same two-photon quantum state. Note that instead of using two identical sets of crystals, one could pass the laser twice through the same set of crystals (see Supplemental Material). Sources $Q_1$ and $Q_2$ emit in such a way that only one pair of photons is produced at a time, i.e., we generate only one copy of the state.  We denote the two photons by $\a$ and $\b$. Suppose that $Q_1$ can emit photon $\a$ into propagation mode $\a_1$. We ensure that $Q_2$ can emit photon $\a$ only in the same propagation mode ($\a_1$). This is done by sending the beam of photon $\a$ generated by $Q_1$ through source $Q_2$ and perfectly aligning the beam with the spatial propagation mode $\a$ generated by $Q_2$. Therefore, if one only observes photon $\a$ that emerges from $Q_2$, one cannot identify the origin of the photon. Stimulated emission at $Q_2$ due to the input of mode $\a_1$ is negligible \cite{wiseman2000induced,lahiri2017can}. Sources $Q_1$ and $Q_2$ can emit photon $\b$ into distinct propagation modes $\b_1$ and $\b_2$, respectively. These two modes are superposed by a beamsplitter, $BS$, and one of the outputs of $BS$ is collected by a detector, $PD$, where the single-photon counting rate (intensity) is measured. We also include an additional device, $\Gamma$, which can transform or project the light emerging from the beamsplitter to a particular state of our choice. Note that photon $\a$ is never detected. It is known that single-photon interference can be observed (at $PD$) for photon $\b$ in such a setup \cite{ZWM-ind-coh-PRL,WZM-ind-coh-PRA}.
\par
We now introduce a device, $O$, in propagation mode $\a_1$ between $Q_1$ and $Q_2$.  The effect of this interaction is observed in the interference pattern recorded at $PD$ although, photon $\b$ never interacts with $O$. Recent imaging, spectroscopy and optical coherence tomography experiments have shown that with the knowledge of the two-photon quantum state, one can retrieve the information about the interaction from the interference pattern \cite{lemos2014,lahiri2015,kulik2016,valles2018optical,miller2019versatile,paterova2019polarization,Paterova_2018,chekhova2016nonlinear}.
\par
Our entanglement verification method is essentially the converse of the imaging method described in Refs. \cite{lemos2014,lahiri2015}. Here, we retrieve the information about the two-photon quantum state from the interference pattern with the knowledge of the interaction between $O$ and the undetected photon $\a$.
\par
In order to demonstrate our method, we work with two-qubit states determined by three free parameters. One example of such state is expressed by the density operator
\begin{align}\label{mixed-state-form}
\dm=&I_H \ket{H_\alpha H_\beta }\bra{H_\alpha H_\beta }
+I_V \ket{V_\alpha V_\beta }\bra{V_\alpha V_\beta }
\nonumber
\\ &+\big(\I\sqrt{I_H I_V}e^{-i\phi}\ket{H_\alpha H_\beta }\bra{V_\alpha V_\beta }+\text{H.c.}\big),
\end{align}
where $I_H+I_V=1$ with $0\leq I_H \leq 1$, $\phi$ is a real number,
and $0\leq\I\leq 1$. This state can be seen as a result of decoherence of the pure state $\sqrt{I_H}\ket{H_\alpha H_\beta}+e^{i\phi}\sqrt{I_V}\ket{V_\alpha V_\beta}$. Note that state $\dm$ can also be obtained generalizing the two following Bell States: $\ket{\Phi^{+}}=(\ket{H_{\a},H_{\b}}+\ket{V_{\a},V_{\b}})/\sqrt{2}$ and $\ket{\Phi^{-}}=(\ket{H_{\a},H_{\b}}-\ket{V_{\a},V_{\b}})/\sqrt{2}$.
Our method also applies to the mixed state which has the form $\dm=I_1 \ket{H_{\a},V_{\b}}\bra{H_{\a},V_{\b}} +I_2
\ket{V_{\a},H_{\b}}\bra{V_{\a},H_{\b}}+(e^{-i\phi}\I\sqrt{I_1 I_2}\ket{H_{\a},V_{\b}}
\bra{V_{\a},H_{\b}} +\text{H.c.})$. It is obtained by generalizing the two Bell states $\ket{\Psi^{+}}=(\ket{H_{\a},V_{\b}}+\ket{V_{\a},H_{\b}})/\sqrt{2}$ and $\ket{\Psi^{-}}=(\ket{H_{\a},V_{\b}}-\ket{V_{\a},H_{\b}})/\sqrt{2}$.
\par
State $\dm$ is entangled when $0<I_H\leq 1$ and $\I \neq 0$. It is maximally entangled for $I_H=I_V=1/2$ and $\I=1$. When $I_H=1$ or $I_H=0$, the state $\dm$ is pure and separable. The state is maximally mixed and separable for $I_H=I_V=1/2$ and $\I=0$. 
A measure of entanglement, commonly used for
two-qubit systems, is the concurrence $\mathcal{C}$
\cite{wootters1998entanglement}, which for the state, $\dm$ [Eq. (\ref{mixed-state-form})], is
\begin{equation}\label{conc-form}
\mathcal{C}(\dm)=2\I\sqrt{I_H I_V}.
\end{equation}
For maximally entangled states $\mathcal{C}(\dm)=1$ and for separable states $\mathcal{C}(\dm)=0$.
\par
\begin{figure}
\centering
 \subfigure[] {
    \label{ent-source}
   \includegraphics[width=0.35\linewidth]{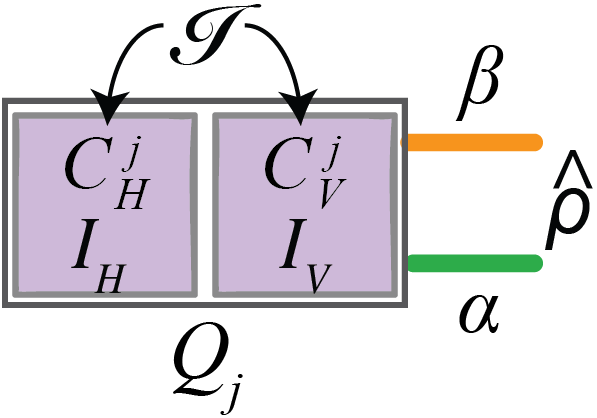}
}
   \subfigure[] {
    \label{3a}
    \includegraphics[width=0.95\linewidth]{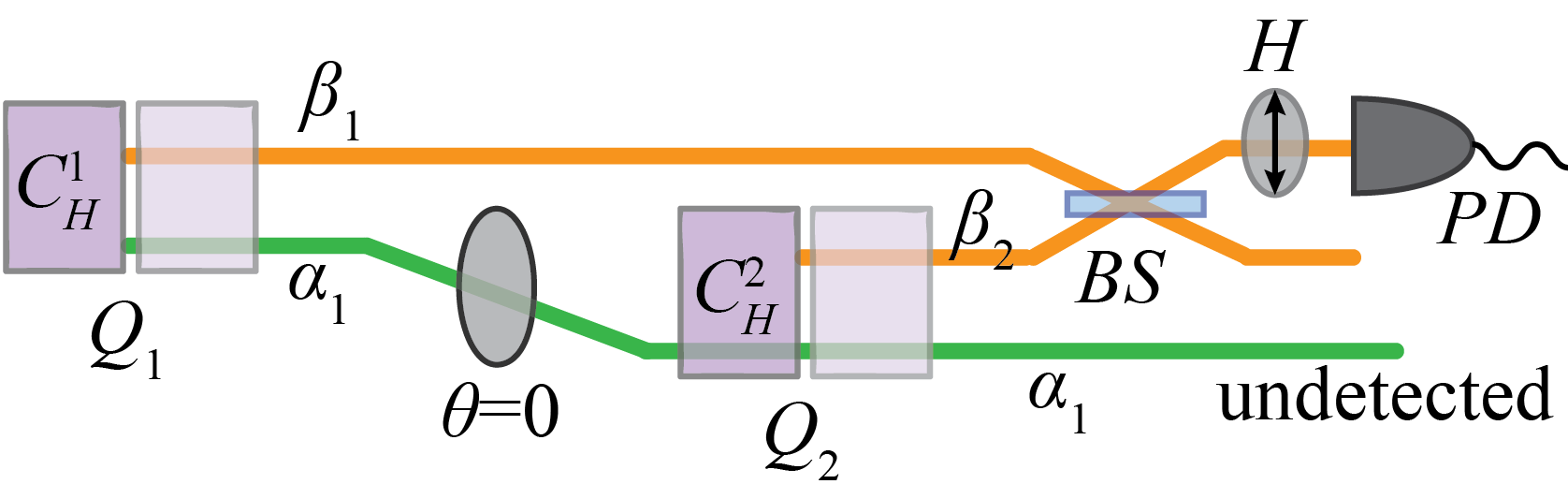}
}
   \subfigure[] {
    \label{3b}
    \includegraphics[width=0.95\linewidth]{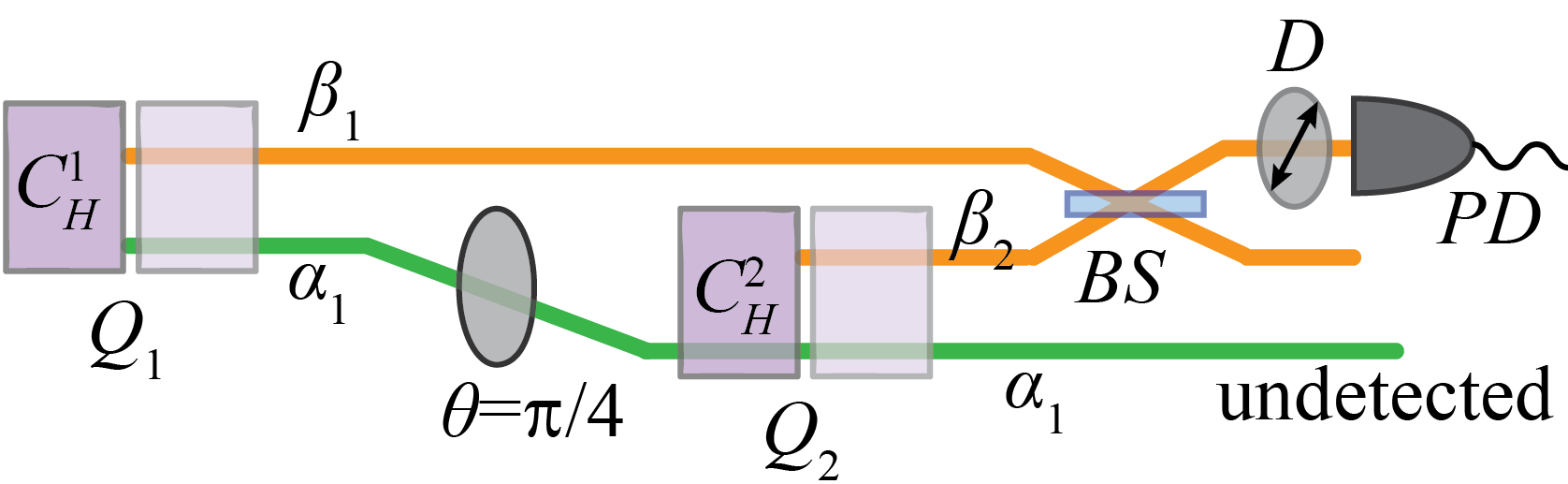}
}
\centering
\caption{(a) Source of the polarization-entangled photon pair: Each source ($Q_j$) is composed of two nonlinear crystals, $C_H^j$ and $C_V^j$, which produce the horizontally polarized ($\ket{H_\alpha H_\beta }$) and vertically polarized ($\ket{V_\alpha V_\beta}$) parts of the entangled state respectively. The relative intensity of
emissions $\ket{H_\alpha H_\beta}$ at $C_H^j$ and $\ket{V_\alpha
V_\beta}$ at $C_V^j$ are $I_H$ and $I_V$ respectively. The
coherence between these two emissions is $\I$. (b) $Q_1$ and $Q_2$ are illuminated by mutually coherent laser propagation modes (not shown) such that the horizontal ($H$) components of the possible emissions at the separate sources are coherent. Highest interference visibility is observed at $PD$ if $H$ polarized photons are detected.
(c) For  $\theta=\pi/4$, we probe the indistinguishability between emissions at $C_H^1$ and $C_V^2$ and also between emissions at $C_V^1$ and $C_H^2$ (not shown) by detecting diagonally ($D$) linearly polarized $\beta$-photons. The visibility of the resulting interference pattern depends on the entanglement in the two-photon state.  }\label{fig3}
\end{figure}

In the experiment our source of entangled photons is a pair of perpendicularly oriented nonlinear crystals, $C_H^j$ and $C_V^j$ (Fig.\ref{ent-source})\cite{KMWZ-pol-ent-source}.
However, our scheme also works for any other source producing the state $\dm$ given by Eq. (\ref{mixed-state-form}), for example, a single type-II non-linear crystal \cite{kwiat1995new}. 
Horizontally and vertically polarized two-photon states
($\ket{H_\alpha H_\beta}$ and $\ket{V_\alpha V_\beta}$) are produced by spontaneous parametric down-conversion in $C_H^j$ and $C_V^j$ respectively. Parameters $I_H$ and $I_V$ are proportional to the probability of emissions at $C_H^j$ and $C_V^j$, respectively. The parameter $\I$ represents the mutual coherence between these emissions and $\phi$ is the relative phase between these emissions. All three parameters are independently tuned in our experiment. 
\par
As illustrated in Fig. \ref{fig:scheme-illus}, we use two such sources in the experiment (see Supplemental Material for the detailed experimental setup). As for device $O$, we use a half-wave plate (HWP), which allows us to introduce distinguishability. The device, $\Gamma$, is a combination of wave plates and a polarizer (Supplemental Information) such that we can project photon $\b$ onto horizontal ($H$), vertical ($V$), diagonal ($D$), anti-diagonal ($A$), right-circular ($R$), and left-circular ($L$) polarization states. Therefore, we choose the measurement basis by the use of $\Gamma$. 
The phase in the interferometer is changed by moving the position of the beamsplitter ($BS$).
\par
The two sources ($Q_1$ and $Q_2$) are illuminated by mutually coherent laser beams. In such a situation photon-pair emissions at $C_H^1$ and $C_H^2$ are fully coherent. If the HWP is set at angle $\theta=0$ and the device $\Gamma$ is set such that only $H$-polarized photons ($\ket{H_\b}$) are detected at $PD$ (Fig. \ref{3a}), visibility of the recorded interference pattern becomes maximum. (This result is fully consistent with the results presented in Refs. \cite{ZWM-ind-coh-PRL,WZM-ind-coh-PRA}.) Note that in this case, no photon emitted by $C_V^1$ and $C_V^2$ arrives at the detector. Likewise, photon-pair emissions at $C_V^1$ and $C_V^2$ are also fully coherent when $Q_1$ and $Q_2$ are illuminated coherently. However, as mentioned before, pair emissions at $C_H^1$ and $C_V^1$ (and also at $C_H^2$ and $C_V^2$) may not be fully coherent and the mutual coherence between them is given by $\I$. If emission at $C_H^1$ are fully coherent to $C_H^2$ and the mutual coherence between emissions at $C_H^2$ and $C_V^2$ is $\I$, then the mutual coherence between pair emissions at $C_H^1$ and $C_V^2$ is also given by $\I$. The same is true for the mutual coherence between emissions at $C_V^1$ and $C_H^2$.
\par
When the HWP is set at $\theta=\pi/4$, the polarization components of $\alpha_1$ are rotated as  $\ket{H_\a}\rightarrow \ket{V_\a}$ and $\ket{V_\a}\rightarrow -\ket{H_\a}$. The quantum state produced at $Q_2$ is not affected by the rotation of the HWP. If we now detect photon $\b$ after projecting onto the $\{\ket{H_\b},\ket{V_\b}\}$ basis, no interference is observed for all values of $\I$ and $I_H$, i.e., the corresponding values of visibility are $\mathcal{V}_{H}\big|_{\theta=\frac{\pi}{4}}=\mathcal{V}_{V}\big|_{\theta=\frac{\pi}{4}}=0$. This is because \textit{if we were} to jointly measure the polarization state of photon $\alpha$ (after $Q_2$) we would know from which crystal photon $\beta$ had arrived. It is important to note that measurement in $\{\ket{H_\b},\ket{V_\b}\}$ basis does not yield any information about entanglement. 
\begin{figure*}
\centering\includegraphics[width=1\linewidth]{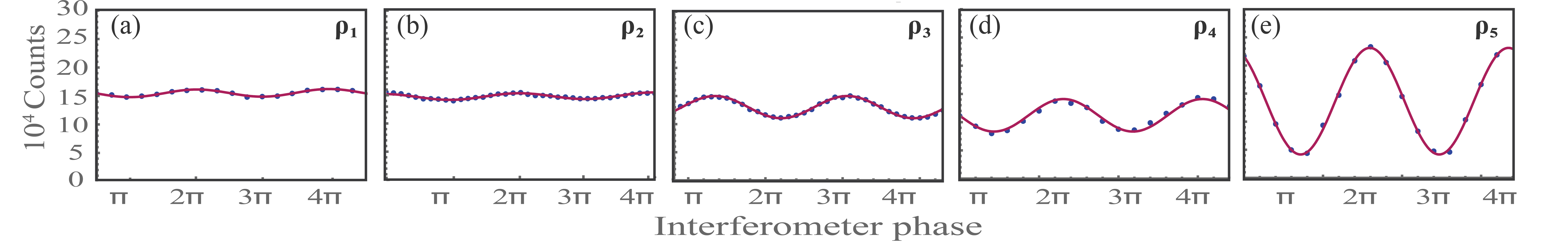}
\caption{Signature of entanglement in single-photon interference visibility. The data show single-photon interference patterns recorded in projective measurements onto state $\ket{D_\b}=(\ket{H_\b}+\ket{V_\b})/\sqrt{2}$ for five mixed states, $\dm$, when HWP angle $\theta=\pi/4$. The interferometer phase is varied using a system of waveplates in the pump. 
The relevant parameters of $\dm$ and the visibilities are: (a)
$I_H=0.96$, $\I=0.25$,  $\mathcal{V}_{D}\big|_{\theta=\frac{\pi}{4}}=0.04$; (b)
$I_H=0.51$,  $\I=0.042$,  $\mathcal{V}_{D}\big|_{\theta=\frac{\pi}{4}}=0.04$; (c) $I_H=0.50$, $\I=0.22$, $\mathcal{V}_{D}\big|_{\theta=\frac{\pi}{4}}=0.15$;
(d) $I_H=0.65$, $\I=0.38$, $\mathcal{V}_{D}\big|_{\theta=\frac{\pi}{4}}=0.26$
(e) $I_H=0.47$, $\I=0.94$, $\mathcal{V}_{D}\big|_{\theta=\frac{\pi}{4}}=0.70$; 
 The lowest visibilities (a and b) correspond to the (almost) separable states ($\rho_1$ and $\rho_2$ in Fig. \ref{Barchart}). The intermediate values of visibilities (c and d) correspond to a  state that is not maximally entangled ($\rho_4$ in Fig. \ref{Barchart}). The highest visibility (e) corresponds to the (almost) maximally entangled state ($\rho_5$ in Fig.\ref{Barchart}). Note that the reduced polarization density matrix for photon $\beta$ in cases (b) and (e) represents the same unpolarized state, yet the single-photon interference visibilities are radically different, allowing us to distinguish between the two cases. The error bars are smaller than the data points in the plots.}  \label{threevis}
\end{figure*}
\par
We now detect photon $\b$ after projecting onto $\ket{D_\b}\equiv (\ket{H_\b} +\ket{V_\b})/\sqrt{2}$ while the HWP is set at $\theta=\pi/4$. Photon $\b$ can now arrive at the detector in four alternative ways: 1) from $C_H^1$, 2) from $C_H^2$, 3) from $C_V^1$, and 4) from $C_V^2$. We first note that alternative 1 is fully distinguishable from alternative 2 for the reason discussed in the previous paragraph. Likewise, alternative 3 is fully distinguishable from alternative 4. For very similar reasons, alternatives 1 \& 3 are also distinguishable from each other, as are alternatives 2 \& 4. 
\par
According to the laws of quantum mechanics, the distinguishable alternatives do not result in interference. Let us now consider the remaining two options: alternatives 1 \& 4, and alternatives 2 \& 3. These two sets of alternatives are fully equivalent to each other. For the sake of brevity, we only present arguments for alternatives 1 \& 4 (Fig. \ref{3b}). We recall that the mutual coherence between emissions at $C_H^1$ and $C_V^2$ is given by $\I$. Therefore, if $\I=0$, alternatives 1 \& 4 become fully distinguishable and no interference occurs \cite{mandel1991coherence}. If $I_H=0$ or $I_V=0$, no emission occurs at $C_H^1$ or $C_V^2$. In this case alternatives 1 \& 4 are also fully distinguishable. When $\I=1$ and $I_H=I_V=1/\sqrt{2}$, alternatives 1 \& 4 are fully indistinguishable and interference occurs with maximum visibility. In any intermediate case interference occurs with reduced visibility. Following this argument, we find that the visibility is given by (c.f. \cite{pol-ent-theory})
\begin{align}\label{vis-D}
\mathcal{V}_{D}\big|_{\theta=\frac{\pi}{4}} \propto \I\sqrt{I_H I_V}.
\end{align}
It follows from Eqs. (\ref{conc-form}) and (\ref{vis-D}) that the single-photon interference visibility $\mathcal{V}_{D}\big|_{\theta=\frac{\pi}{4}}$ is linearly proportional to the concurrence, i.e., the visibility contains information about the entanglement. Figure \ref{threevis} shows experimentally obtained interference patterns for four sates. The data clearly show that when HWP angle $\theta=\pi/4$, the visibility measured for $\ket{D_\b}$ increases with the amount of entanglement . 
\par
Analogous arguments apply to measurements in the circular polarization basis $\{\ket{R_\b},\ket{L_\b}\}$. Hence,   non-zero visibility after projecting photon $\b$ onto $\ket{D_\b}$, $\ket{A_\b}$, $\ket{R_\b}$ or $\ket{L_\b}$ confirms that the two-photon state is entangled.
\begin{figure}[htbp]
\centering\includegraphics[width=1\linewidth]{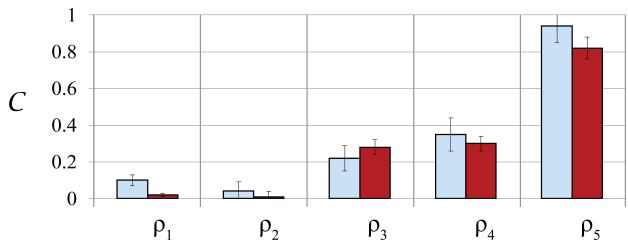}
\caption{Experimentally measured concurrence. The blue bars show the results, obtained by our scheme where only subsystem $\beta$ is detected (singles counts). In order to generate the five states $\dm_1,\dots,\dm_5$, the parameters $I_H$ and $\I$ (Eq.\ref{mixed-state-form}) were varied (Supplemental Information). We compare our results with the values of concurrence obtained from the full two-qubit
tomography (red bars). State $\dm_1$ is  approximately pure and separable; state $\dm_2$ is almost maximally mixed (separable); states $\dm_3$ and $\dm_4$ are mixed and entangled; state $\dm_5$ is almost pure and maximally entangled.}\label{Barchart}
\end{figure}
\par
Equations (\ref{conc-form}) and (\ref{vis-D}) suggest that the concurrence can be determined from the visibility of the interference patterns. However, for an accurate measurement of the concurrence one needs to consider the experimental loss of photons in propagation mode $\a_1$ between $Q_1$ and $Q_2$ and source distinguishability because these lead to reduction of visibility. In fact, visibilities measured for $\ket{D_\b}$, $\ket{A_\b}$ and $\ket{R_\b}$, and $\ket{L_\b}$ (while $\theta=\pi/4$) will always be smaller or equal to $\mathcal{C}(\dm)/\sqrt{2}$.  Since losses for $H$ and $V$ polarization components are different in our experiment, we need to calibrate the system by measuring single-photon visibility in $\{\ket{H_\b},\ket{V_\b}\}$ basis for $\theta=0$. We find that the concurrence is given by \cite{pol-ent-theory}
\begin{align}\label{conc-vis}
\mathcal{C}(\dm)=\sqrt{2
\frac{\left(\mathcal{V}_{D}\big|_{\theta=\frac{\pi}{4}}\right)^2
+\left(\mathcal{V}_{R}\big|_{\theta=\frac{\pi}{4}}\right)^2}
{\left(\mathcal{V}_{H}\big|_{\theta=0}\right)^2
+\left(\mathcal{V}_{V}\big|_{\theta=0}\right)^2}}.
\end{align}
\par
It is important to note that the denominator in the equation above is the calibration of our interferometer. In the ideal case, one wouldn't have to measure the interference visibilities for  $\theta=0$, because they would simply be equal to one. However, in general, interferometric visibility is not maximum even in that case, due to path/source distinguishability introduced by degrees of freedom other than polarization, for example, losses inside the interferometer \cite{ZWM-ind-coh-PRL,WZM-ind-coh-PRA}, imperfect path adjustment \cite{zou1993control}, differences between the two sources, and beam propagation between the crystals \cite{barbosa1993degree, grayson1994spatial}. Including the numerator in Eq.\ref{conc-vis} guarantees that our entanglement measure is robust to alignment, loss and path length imperfections, among others. As long as the interference for $\theta=0$ is larger than zero, our method can be applied.
\par 
Notice also that if one doesn't do the calibration with $\theta=0$, one can nevertheless use our method as an entanglement witness. In other words,  $\left(\mathcal{V}_{D}\big|_{\theta=\frac{\pi}{4}}\right)^2
+\left(\mathcal{V}_{R}\big|_{\theta=\frac{\pi}{4}}\right)^2>0$ implies that the biphoton state is entangled in polarization.
\par
The experimentally measured values of concurrence for five mixed states, $\dm_1,\dots,\dm_5$, are shown in Fig. \ref{Barchart}. For comparison, we also make tomographic reconstruction of these states (Supplemental Material) and determine the concurrences independently. As can be clearly seen in Fig. \ref{Barchart}, the values of concurrence obtained by our method (without coincidence detection) are in excellent agreement with those values obtained from quantum state tomography (with coincidence detection). 
\par
Note that by measuring the relative horizontally polarized and vertically polarized photon count rates produced by \textit{one single source} ($Q_1$ or $Q_2$), one can obtain the parameter $I_H$. One can then use the value of $C(\dm)$, obtained from the single photon interference visibilities, to determine the parameter $\I$. The corresponding results are in very good agreement with those obtained from full quantum state tomography (Supplemental Information).

\par
In summary, we have verified and measured entanglement in bipartite mixed states without detecting one subsystem. Our method is particularly useful when, for any reason, detectors are not available for one of the subsystems. The method is resistant to experimental imperfections, such as alignment imperfections, loss, temporal walk-off, spectral differences, path length imperfections, and non-identical sources. These imperfections reduce visibility and are quantified by the denominator on the right-hand side of Eq.~(\ref{conc-vis}). In fact, Eq.~(\ref{conc-vis}) gives the appropriately calibrated concurrence for the bipartite state considered here. Detailed instructions on how to analyze, how to model theoretically, and how to maximize interference visibility in such interferometers can be found in Ref.~\cite{BarretoLemos:22}.
\par
It is important to note that the method is independent of the structure of each source, \textit{these need not be composed of two crystals}. In addition, there is no need for two identical sources, as a double pass of the laser in the same source would work (Supplemental Material). We demonstrated the method by working with a mixed state that is obtained by generalizing two Bell states. Our method also applies to the mixed state which can be obtained by generalizing the other two Bell states \footnote{This mixed state has the form $\dm=I_1 \ket{H_{\a},V_{\b}}\bra{H_{\a},V_{\b}} +I_2
\ket{V_{\a},H_{\b}}\bra{V_{\a},H_{\b}}+(e^{-i\phi}\I\sqrt{I_1 I_2}\ket{H_{\a},V_{\b}}
\bra{V_{\a},H_{\b}} +\text{H.c.})$. It is obtained by generalizing the two Bell states $\ket{\Psi^{+}}=(\ket{H_{\a},V_{\b}}+\ket{V_{\a},H_{\b}})/\sqrt{2}$ and $\ket{\Psi^{-}}=(\ket{H_{\a},V_{\b}}-\ket{V_{\a},H_{\b}})/\sqrt{2}$.}. Furthermore, the method could also be extended to transverse spatial entanglement \cite{hochrainer2017quantifying,lahiri2017twin} or orbital angular momentum entanglement, if devices $O$ and $\Gamma$ (Fig. \ref{fig:scheme-illus}) are appropriately chosen. Although we demonstrated the method using photonic states, the principle underlying the method is applicable to other quantum systems.

\par
\textbf{Acknowledgements.} The experiment was performed at the Institute for Quantum Optics and Quantum Information,
Austrian Academy of Sciences, Boltzmanngasse 3, Vienna A-1090, Austria. We acknowledge support from the Austrian Academy of Sciences (\"OAW- 462 IQOQI, Vienna) and the Austrian Science Fund (FWF) with SFB F40 (FOQUS) and W1210-2 (CoQus). G.B.L. also acknowledges support from the Brazilian National Council for Scientific and Technological Development (CNPq),  from the Fundação de Amparo à Pesquisa do Estado do Rio de Janeiro - FAPERJ, and from the Coordenação de
Aperfeiçoamento de Pessoal de Nível Superior (CAPES - Brasil) -- Finance Code 001. M.L.  acknowledges support from College of Arts and Sciences and the Office of the Vice President for Research, Oklahoma State University. R.L. was supported by a National Science Centre (Poland) grant 2015/17/D/ST2/03471 and the Foundation for Polish Science under the FIRST TEAM project 'Spatiotemporal photon correlation measurements for quantum metrology and super-resolution microscopy' co-financed by the European Union under the European Regional Development Fund (POIR.04.04.00-00-3004/17-00).


%

\end{document}